# Convolutional Image Edge Detection Using Ultrafast Photonic Spiking VCSEL-Neurons

J. Robertson, Y. Zhang, M. Hejda, A. Adair, J. Bueno, S. Xiang, A. Hurtado

*Abstract*—We report experimentally and in theory on the detection of edge information in digital images using ultrafast spiking optical artificial neurons towards convolutional neural networks (CNNs). In tandem with traditional convolution techniques, a photonic neuron model based on a Vertical-Cavity Surface Emitting Laser (VCSEL) is implemented experimentally to threshold and activate fast spiking responses upon the detection of target edge features in digital images. Edges of different directionalities are detected using individual kernel operators and complete image edge detection is achieved using gradient magnitude. Importantly, the neuromorphic (brain-like) image edge detection system of this work uses commercially sourced VCSELs exhibiting spiking responses at sub-nanosecond rates (many orders of magnitude faster than biological neurons) and operating at the telecom wavelength of 1300 nm; hence making our approach compatible with optical communication and data-center technologies. These results therefore have exciting prospects for ultrafast photonic implementations of neural networks towards computer vision and decision making systems for future artificial intelligence applications.

*Index Terms*—Neuromorphic Photonics, VCSELs, Nonlinear Dynamics, Convolutional Neuronal Networks, Computer vision.

## I. Introduction

TRAINING computers to interpret and recognise images, more commonly known as computer vision, is a complex task being undertaken using a combination of image processing and machine learning techniques [1]. Branching from artificial intelligence (AI), computer vision is now seeing a surge in interest as the goal to automate traditionally human tasks becomes more prevalent with the rapid development of fields such as medical image analysis [2] and autonomous vehicles [3]. One computing architecture that has shown promise in similar branches of AI, such as pattern and voice recognition [4-5], and data classification [6], is artificial neural networks (ANNs) [7]. Based on the vast interconnected networks of biological neurons in the brain, ANNs, and their massive parallelism, are capable of performing complex, human-like reasoning better than traditional computing architectures. A type of ANN that focus prominently on image recognition tasks and computer vision is Convolutional Neural Networks (CNNs). CNNs apply the traditional image processing technique, convolution, in a large parallel network and achieve overall recognition by assembling and comparing many smaller features [8]. Hence, feature extraction, searching for interest points, lines, edges and corners, is a key fundamental process that occurs in the layers of CNNs [9].

However, due to the large number of convolution operations and the complicated interconnected architecture, CNNs suffer from high power and computational resource requirements. Many electronic CNNs utilize dedicated hardware or additional GPUs to operate [10-12], restricting the footprint and application of these systems that are already contested by the near-fundamental limit of electronic technology [13]. One solution that could help alleviate these undesirable requirements would be a change to a photonic-based platform. Due to the bosonic nature of photons, neural networks based upon photonic techniques hold a number of benefits over their electronic counterparts, such as low power requirements, large bandwidth, increased operation speed and low cross-talk.

Recent experimental and numerical reports on photonic-based CNNs have emerged proposing orders of magnitude improvements to operational speeds using systems based upon silicon weighting banks [14] and modulator arrays [15-17]. Similarly, Semiconductor Lasers (SLs) are highly promising photonic devices for ANN building blocks (see [18] for a review). SL models based on vertical-cavity surface-emitting lasers (VCSELs) specifically have shown the capability to perform neuromorphic actions similar to those in observed in biological neurons [19-30]. Key behaviours such as tonic spiking [21], spike thresholding and inhibition [22-23], interconnectivity [24-26,28], input integration, etc., have all been demonstrated in these compact SLs at sub-nanosecond spiking rates. Also, computing functionalities such as logic and pattern recognition have recently been achieved using the

Manuscript received May 26, 2020. This work was supported in part by the Office of Naval Research Global under Grant ONRGNICOP-N62909-18-1-2027, the European Commission under Grant 828841-ChipAI-H2020-FETOPEN2018-2020, the UK's EPSRC Doctoral Training Partnership (EP/N509760) and the National Natural Science Foundation of China (No. 61974177, No.61674119).

J. Robertson, M. Hejda, A. Adair, J. Bueno and A. Hurtado are with the Institute of Photonics, University of Strathclyde, Technology and Innovation Centre, 99 George St., G1 1RD, Glasgow, UK (phone: +44 (0)141 548 3668; e-mail: joshua.robertson@strath.ac.uk; antonio.hurtado@strath.ac.uk).

S. Xiang is with State Key Laboratory of Integrated Service Networks, Xidian University, Xi'an 710071, China (e-mail: jxxsy@126.com).

Y. Zhang is with the Institute of Photonics, University of Strathclyde, Technology and Innovation Centre, 99 George St., G1 1RD, Glasgow, UK and with State Key Laboratory of Integrated Service Networks, Xidian University, Xi'an 710071, China.

sub-nanosecond spiking representation [29,30]. Furthermore, theoretical studies of these devices, based on the spin-flip model (SFM) [31,32], also suggest their capability to perform learning based upon spike-timing dependent plasticity (STDP) as observed in biological neural networks [33-34]. In tandem with their impressive neuronal capabilities, VCSELs possess key beneficial properties: they operate at ideal telecom wavelengths, have reduced manufacturing costs, and have compact, integrable structures. VCSELs, and the fast (sub-nanosecond) spiking dynamics that they yield, make enticing candidates for SL-based photonic implementations of CNNs for image recognition and computer vision.

In this work, we propose and demonstrate an artificial spiking VCSEL-neuron for use in primitive feature-extracting CNN layers. We provide both experimental and theoretical results (based on the SFM [31,32], for the edge detection of digital images at very high speed using a single photonic spiking VCSEL-neuron operating at the key telecom wavelength of 1300 nm. This paper is organised as follows: in sections II & III we discuss the convolutional and experimental technique applied to achieve image edge-detection at ultrafast rates with a spiking photonic VCSEL-neuron. Section IV describes the theoretical model used to validate the experimental findings. Section V provides theoretical and experimental results on vertical and horizontal image edge detection with the VCSEL-based photonic spiking neuron. In section VI we discuss and provide results of image gradient magnitude detection.

## II. IMAGE PROCESSING AND CONVOLUTION TECHNIQUE FOR IMAGE EDGE DETECTION

In this work, image convolution is incorporated alongside an experimental realisation of a photonic spiking VCSEL-neuron (via modulated optical injection), as shown in Fig. 1. In this section, we detail the image processing and image convolution technique applied (Fig. 1a). Initially, a pre-processing stage performs the conversion of digital greyscale source images to positive (black) and negative (white) integer (1 and -1) matrices. Source images that contain different directional features were selected as shown in Fig. 2. The 28x28 pixel cross and saltire (Scotland's national flag) images (Fig. 2a-b) were chosen as they contain horizontal, vertical and diagonal features. The larger 50x50 pixel image in Fig. 2c corresponds to the logo of our institute, the Institute of Photonics (IOP) at the University of Strathclyde. The IOP logo contains additional curved features and is much larger in size to illustrate the versatility of our work's technique. In order to reveal edge information in these images convolution is performed using a 2x2 kernel operator. The latter applies a weight to each pixel in a 2x2 region of the image and sums of all 4 weighted pixel values. The 4-value sum corresponds to the destination pixel value in the new convolved image (as shown in Fig. 1a). Different features can be targeted for recognition by applying different kernel operators and 2x2 kernels are scanned along every pixel in a row, and every row in an image. By comparing neighbouring pixels in this convolution process, we are able to identify features that best match our selected kernel operator. This process can be summarised using the following equation:

$$g_{p,q} = \sum_{m=0}^{M} \sum_{n=0}^{N} f_{p+m,q+n} \cdot K_{m,n} \qquad (1)$$

where $g_{p,q}$ is the value of the destination pixel when the source image anchor-pixel $f_{p,q}$ is operated on by kernel $K$. A (M+1) x (N+1) pixel neighbourhood is operated on by the customisable kernel operator, in this work we set M = N = 1 to achieve a 2x2 neighbourhood array. No image buffer was used hence the dimensionality of the new convolved image is reduced by 1.

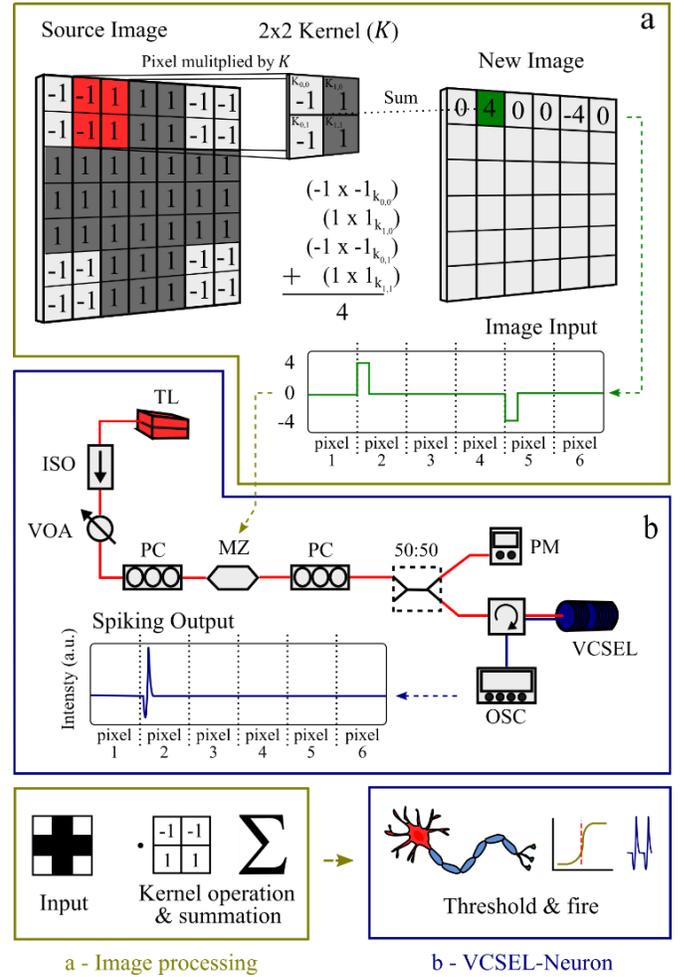

Fig 1. Image convolution technique utilised to obtain high-speed spiking image edge detection and the experimental setup employed to implement it. The image processing procedure (a) and the experimental realisation of the VCSEL-neuron (b). In (a), a black and white source image is converted into positive (1) and negative (-1) integers before it is multiplied by a 2x2 kernel operator. The resulting image is converted into a RZ image input where the destination pixel value is taken and encoded into the tunable laser's optical intensity. In (b), light from a tunable laser is encoded with the convolved image input using a Mach Zehnder intensity modulator (MZ). The intensity encoded signal is injected into the spiking VCSEL-neuron whose response is collected via the optical circulator and analysed using a fast real-time oscilloscope. Two polarisation controllers (PC), an optical isolator (OI) and a variable optical attenuator (VOA) are used to control the light signals within the fibre-optic based experimental setup of this work.

In this work the VCSEL-neuron (Fig. 1b) will threshold the destination pixel value $g_{p,q}$ emitting ultrashort (sub-nanosecond) neuron-like spikes to determine which pixels in the original source image contain the target feature. In order to generate an input signal that can be easily injected into our VCSEL-neuron, time-division multiplexing was used. In the time-multiplexed image input, each destination pixel was sequentially allocated the same configurable pixel duration. The latter was selected equal to 1.5 ns/pixel to coincide with the spiking refractory period of the VCSEL-neuron [29]. This, along with an inter-pixel return-to-zero (RZ) coding scheme, helped to ensure all neighbouring pixels were capable of triggering a single spiking event per input. All post-kernel destination pixel values were held for 0.25 ns before returning to zero. The time-division multiplexed image input was produced in an arbitrary waveform generator (AWG) and fed to the experimental setup (shown in Fig. 1b).

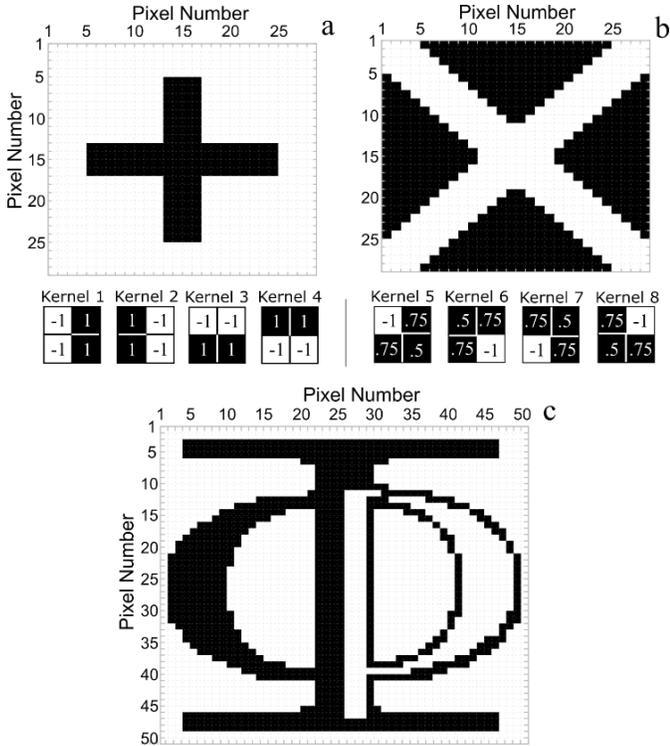

Fig 2. Black and white source images of a cross (a), a saltire (Scotland's national flag) (b) and the logo of the Strathclyde's Institute of Photonics (IOP) (c). The vertical (Kernels 1-2), horizontal (Kernels 3-4) and diagonal (Kernels 5-8) operators are used to detect specific edge features in the source images. Source images (a) and (b) have a resolution of 28x28 pixels whereas source image (c) has a larger resolution of 50x50 pixels.

### III. EXPERIMENTAL VCSEL-NEURON IMPLEMENTATION FOR ULTRAFAST SPIKING EDGE DETECTION

The experimental setup used to implement the photonic VCSEL-neuron and perform spiking image edge detection is shown in Fig. 1b. Similar to our previous work [28], modulated optical injection was used to trigger ultrafast spiking responses from the VCSEL-neuron. Light from a 1300 nm tunable laser (TL) was optically encoded with the time-division multiplexed image input (from the initial image processing and image convolution step). A Mach Zehnder intensity modulator was configured to introduce drops of intensity in the tunable laser's light when subject to positive pulses from the image input. The encoded optical signal was passed to a coupler where a power meter (PM) made a measure of input power, and a circulator injected the optical signal into the VCSEL-neuron. An amplified 9 GHz bandwidth photodetector was used to collect the output of the VCSEL-neuron and a high-speed 13.5 GHz bandwidth and 40 GSa/s sampling rate real-time oscilloscope was used for temporal analysis. Throughout this work, the commercially sourced, fibre-pigtailed VCSEL device was biased with a current of 6.5 mA ($I_{th}$ = 2.96 mA) and temperature stabilised at 298 K. Under these operating conditions, the VCSEL exhibited single mode lasing and the presence of two linear-orthogonally polarised modes, namely a main lasing (parallel) and a subsidiary-attenuated (orthogonal) polarisation mode. Optical injection was made at a frequency detuning ($\Delta f$) of -4.58 GHz from the peak of the subsidiary (orthogonal) mode, inducing polarisation switching during injection locking [29]. An injection power of 152.7 µW was used to injection lock the device. Encoded intensity drops of sufficient amplitude were used to force the laser out of injection locking and into a regime of fast spiking dynamics. Input thresholding is therefore performed by the VCSEL-neuron, allowing the system to reveal target feature information through the triggering of fast neuromorphic spiking events.

### IV. THEORETICAL ANALYSIS OF THE VCSEL-NEURON WITH THE SPIN FLIP MODEL (SFM)

We use a modified version of the well-known spin-flip model (SFM) [32] to evaluate theoretically the operation of the VCSEL-neuron and validate the experimental findings. The modified model used here includes an additional term that accounts for the convolution step, namely the injection of the time varying post-kernel image input. The modified rate equations are as shown below:

$$\frac{dE_{x,y}}{dt} = -(k \pm \gamma_a)E_{x,y} - i(k\alpha \pm \gamma_p)E_{x,y} \\ + k(1+i\alpha)(NE_{x,y} \pm inE_{x,y}) \quad (2) \\ + k_{inj}E_{inj}(t)e^{i\Delta\omega_x t} + F_{x,y}$$

$$\frac{dN}{dt} = -\gamma_N[N(1+|E_x|^2+|E_y|^2) - \mu \\ + in(E_y E_x^* - E_x E_y^*)] \quad (3)$$

$$\frac{dn}{dt} = -\gamma_s n - \gamma_N[n(|E_x|^2+|E_y|^2) \\ + iN(E_y E_x^* - E_x E_y^*)] \quad (4)$$

where the subsidiary (orthogonally-polarised) and solitary (parallel-polarised) lasing modes of the VCSEL are represented by subscripts x and y respectively. The field amplitudes of the subsidiary and solitary modes are represented by $E_x$ and $E_y$. $N$

is the total carrier inversion between conduction and valence bands and $n$ is the carrier inversion difference between spins of opposite polarity. $\gamma_a$ is the gain anisotropy (dichroism) rate, $\gamma_p$ is the linear birefringence rate, $\gamma_n$ is the decay rate of the carrier inversion and $\gamma_s$ is the spin-flip rate. $k$ is the field decay rate, $\alpha$ is the linewidth enhancement rate and $\mu$ is the normalized pump current ($\mu = 1$ represents the VCSEL's threshold). $E_{inj}$ represents the post-kernel image input created during the convolution process and $k_{inj}$ is the injection strength. The angular frequency detuning is defined as $\Delta\omega_x = \omega_{inj} - \omega_0$, where the central frequency $\omega_0 = (\omega_x + \omega_y)/2$ lies between the frequencies of the subsidiary $\omega_x = \omega_0 + \alpha\gamma_a - \gamma_p$ and the solitary mode $\omega_y = \omega_0 + \gamma_p - \alpha\gamma_a$. $\Delta f = f_{inj} - f_x$ is the frequency detuning between the injected field and the subsidiary mode, hence $\Delta\omega_x = 2\pi\Delta f + \alpha\gamma_a - \gamma_p$. The spontaneous emission noise $F_x$ and $F_y$ are calculated as:

$$F_x = \sqrt{\frac{\beta_{sp}\gamma_n}{2}}(\sqrt{N+n}\xi_1 + \sqrt{N-n}\xi_2) \quad (5)$$

$$F_y = -i\sqrt{\frac{\beta_{sp}\gamma_n}{2}}(\sqrt{N+n}\xi_1 - \sqrt{N-n}\xi_2) \quad (6)$$

where $\beta_{sp}$ represents the spontaneous emission strength and $\xi_{1,2}$ represent two independent Gaussian white noise terms of zero mean and a unit variance. The model was solved using the fourth order Runge-Kutta method and the following parameters: $\gamma_a = 2$ ns$^{-1}$, $\gamma_p = 128$ ns$^{-1}$, $\gamma_n = 0.5$ ns$^{-1}$, $\gamma_s = 110$ ns$^{-1}$, $\alpha = 2$, $k = 185$ ns$^{-1}$, $k_{inj} = 15$ ns$^{-1}$ and $\beta_{sp} = 10^{-5}$.

## V. VERTICAL AND HORIZONTAL EDGE DETECTION IN SOURCE IMAGES USING A SPIKING VCSEL-NEURON

The detection of horizontal and vertical edge features was first tested using the 'cross' source image in Fig. 2a. Kernels 1-4 (shown in Fig. 2) were sequentially applied to the source image and the resulting input values were experimentally injected into the VCSEL-neuron. Kernels 1-2 target vertical lines that transition from white-to-black (Kernel 1) and black-to-white (Kernel 2) pixels. Kernels 3-4 target in turn horizontal lines that transition from white-to-black (Kernel 3) and black-to-white (Kernel 4) pixels respectively. Figs. 3(a-b) and 3(e-f) show respectively the spiking responses measured at the output of the VCSEL-neuron when applying individually Kernels 1-4. Image reconstruction maps are built to depict the collected time series from the VCSEL-neuron as intensity colour maps, where the spiking responses appear yellow and the resting state appears blue. Pixels with spiking responses should indicate the presence of the target feature in the source image. As expected, Figs. 3a and 3b demonstrate the triggering of a spike in response to vertical edges in the cross image. In Fig. 3c the post-kernel image input, created using vertical Kernel 1, is shown for row 10 of Fig. 3a. The image input injects a positive pulse for the detection of a matching target feature and a negative pulse for the detection of an inverse target feature. The corresponding VCSEL-neuron's response, shown is Fig. 3d, demonstrates that only the positive pulse triggers a fast ~100 ps spike at the output of the VCSEL-neuron, highlighting the detection of the target feature. Therefore, applying Kernel 2 as shown in Fig. 3b successfully detects the opposite vertical edges in our image.

Similarly, just as Figs. 3a and 3b demonstrate vertical edge detection in the injected 'cross' patterned source image, Figs. 3e and 3f demonstrate in turn horizontal edge detection. The image input and system response for row 12 of Fig. 3e (Kernel 3) are shown in the time series of Figs. 3g and 3h. Here, as Kernel 3 is scanned horizontally along that specific row of the source image, we create an input consisting of multiple positive target detections. The VCSEL-neuron responds to this input by firing multiple spiking events, one for each of the target detections. The spiking system does not trigger off the half-amplitude pulses (corresponding to corner edges) as the encoded input energy was not enough to cross the spiking activation threshold of the device. These results demonstrate that the experimental spiking system can detect both horizontal edges when applying Kernels 3 and 4.

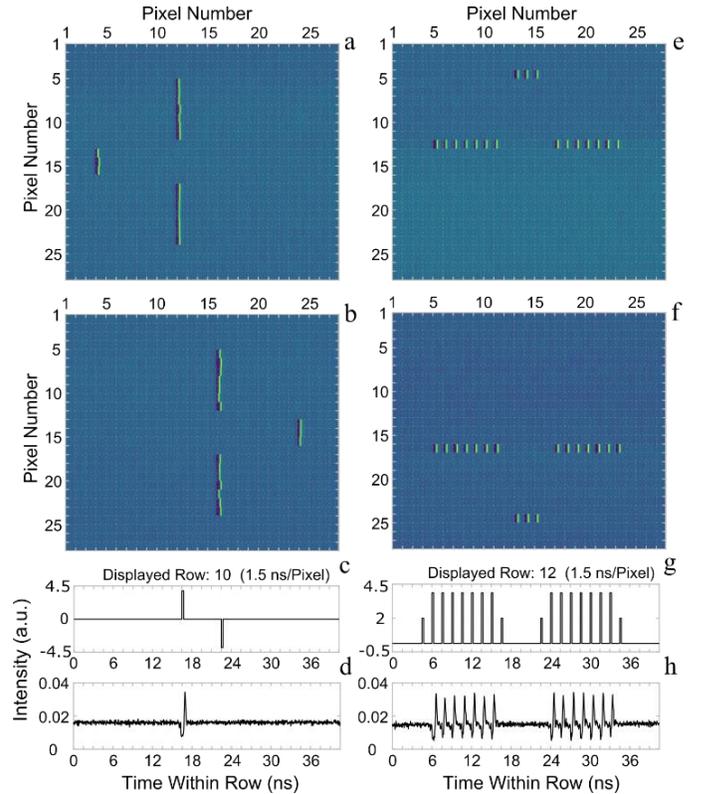

Fig 3. Images built from the spiking responses obtained at the output of the VCSEL-neuron when Kernels 1 (a), 2 (b), 3 (e) and 4 (f) are applied to the cross image and injected into the system. The image input and output time series, corresponding to the selected rows in (a) and (e), are plotted at the bottom of the figure. Input (c) and output (d) correspond to row 10 of (a). Input (g) and output (h) correspond to row 12 of (e). Pixel duration is set to 1.5 ns/pixel in all cases.

For comparison, Fig. 4 shows the theoretically calculated response of the VCSEL-neuron (using the model described in Section IV) when the same 'cross' patterned image is injected into the system model. From the calculated maps reconstructing the image (plotted in white and blue for distinction with the experimental findings of Fig. 3) we see excellent agreement with the measured results. The same number of spiking responses are activated when each kernel is applied to detect different vertical and horizontal image edge features. Also, a similar spiking threshold is achieved preventing the activation of corner edges. The spiking rate achieved in the theoretical results also showed a good correlation with the experiment allowing the model to operate at 1.5 ns/pixel. Therefore, both experiment and theory agree that Kernels 1-4 can be used to successfully perform the spiking edge detection of vertical and horizontal lines from source images using an artificial spiking VCSEL-neuron. Additionally, we found that in both theory and experiment, the spiking threshold could be controlled by varying injection power and frequency detuning, and that it could grant the detection of non-target features with smaller amplitude inputs (such as corners).

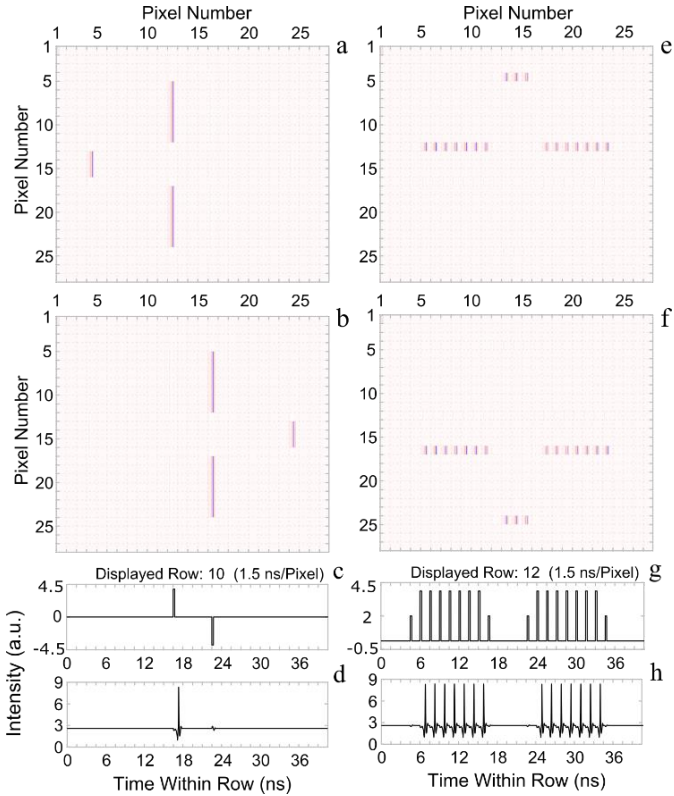

Fig 4. Simulations of the spiking response from the VCSEL-neuron, using the spin-flip model, when Kernels 1 (a), 2 (b), 3 (e) and 4 (f) are applied to the source 'cross' image before its injection into the system. Similar to Fig. 3, inputs and outputs are plotted for row 10 of (a) and row 12 of (e).

## VI. GRADIENT-BASED EDGE DETECTION IN SOURCE IMAGES WITH A SPIKING VCSEL-NEURON

In order to progress beyond the detection of one individual directional edge per input, we look towards performing gradient edge detection. Gradient is a vector with both a direction and a magnitude, which provide information about the rate of change of pixel intensity. Specifically calculating the magnitude of the gradient creates a set of data that can be injected into a VCSEL-neuron to perform spiking edge detection. The magnitude of the gradient $|G(x,y)|$ can be calculated using the following equation:

$$|G(x,y)| = \sqrt{G_x^2 + G_y^2} \qquad (7)$$

where $G_x$ and $G_y$ are respectively the result of convolving a horizontal and a vertical kernel with the image. Both horizontal and vertical kernels must be 90° rotations of one another. By combining the results of two kernel operators in this way, we can detect edges in our images indiscriminately of their direction. Consequently, the convolution results of Kernels 1 and 3 were combined in this way to produce the gradient magnitude. The latter was taken in place of the pixel value when creating the post-kernel image input for injection into the VCSEL-neuron. Gradient-based edge detection was performed on all three source images included in Fig. 2 and the results for both experimental and theoretical approaches are showcased in Figs. 5 and 6.

Figs. 5a and 5c show the experimental response from the VCSEL-neuron when gradient-based edge detection is performed on the 'cross' source image of Fig. 2a. The time series demonstrates that we now have a spiking response for both vertical edges (white-to-black and black-to-white transitions) and the image maps, built from the VCSEL-neuron's spiking output, reveal also the successful detection of the horizontal edges in the 'cross'. Figs. 5b and 5d also demonstrate the activation of spiking events for each edge of the 'cross' image showing excellent agreement with the experimental results. The experimental gradient edge detection of the 'Saltire' image (Fig. 2b) is shown in Fig. 5e illustrating also the successful detection of diagonal edges from a source image with our VCSEL-neuron. Despite gradient detection combining a horizontal and a vertical kernel, we see from the image map (Fig. 5e) and time series (Fig. 5g) that all diagonal edges were successfully detected in the 'Saltire' source image. The modelling of the VCSEL-neuron's response provided in Figs. 5f and 5h again shows excellent agreement with the experimental results, revealing the successful detection of all the diagonal edges. Fig. 6 shows the experimental and theoretical response of the VCSEL-neuron to the gradient edge detection of the 'IOP' logo source image included in Fig. 2c. This larger 50x50 pixel image contained both straight and curved lines (as shown in Fig. 2c), distributed unevenly across the image background. The image map and time series reveal that the gradient detection successfully reveals every directionality of edge. Despite the larger image, the pixel duration remained consistent and only a larger overall time series was required. Again, the modelling of the VCSEL-neuron (Figs. 6b and 6d) showed excellent agreement with the experimental results. Spiking edge detection can therefore be successfully performed by calculating the magnitude of the gradient and using the VCSEL-neuron to threshold the convolved inputs.

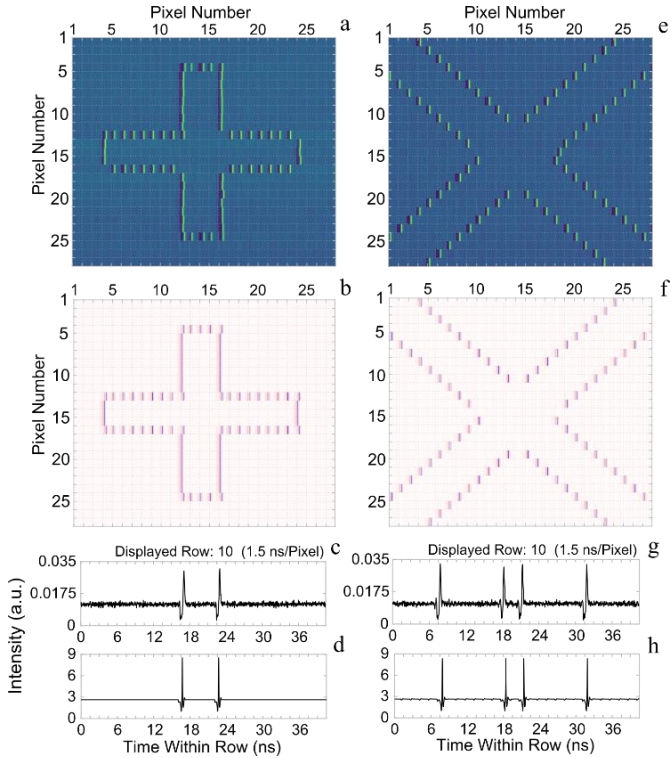

Fig 5. Experimental (blue) and theoretical (white) responses when Kernel 1 and 3 are combined to detect gradient magnitude in the cross (a-d) and saltire (e-h) source images. (a,e) & (b,f) plot respectively experimental and theoretical image reconstruction maps obtained from the spiking responses at the VCSEL-neuron's output. (c) & (d) plot respectfully the experimental and theoretical time series of row 10 in (a) and (b). (g) and (h) plot respectively the experimental and theoretical time series of row 10 in (e) and (f).

## VII. Conclusion

In summary, we demonstrate for the first time to our knowledge, a neuromorphic photonic system based on a VCSEL-neuron performing spiking image edge detection at ultrafast speed, wielding short (<100 ps long) spikes for operation. The artificial neuronal model presented demonstrates the spiking edge detection of vertical, horizontal and diagonal straight lines using individual 2x2 kernel operators and traditional image convolution. Building upon this, the identification and extraction of multiple straight and curved edges, irrespective of their directionality, was achieved in a single input run by calculating and thresholding image gradient magnitude. The system also demonstrates that the spiking threshold could be controlled using system parameters such as injection power and frequency detuning to reveal additional feature information. Furthermore, the experimental findings in this work were shown to have excellent agreement with numerical simulations carried out using a modified version of the SFM. The ultrafast image input, here demonstrated at 1.5 ns/pixel, was selected because of the refractory period of the activated spiking dynamics in the VCSEL-neuron. Yet, we are confident that sub-nanosecond long pixel inputs (GHz operation) could be achieved by additional device design optimisation stages, beyond the scope of this study. The commercially available, compact and telecommunication compatible photonic VCSEL-neurons of this work therefore have high prospects for ultrafast feature-extracting CNN layer implementation. We believe that this VCSEL-neuron system demonstrates powerful neuromorphic functionalities, and that beyond feature extraction, has the potential to enable full implementations of ultrafast photonic ANNs capable of computer vision, pattern recognition and other complex processing tasks.

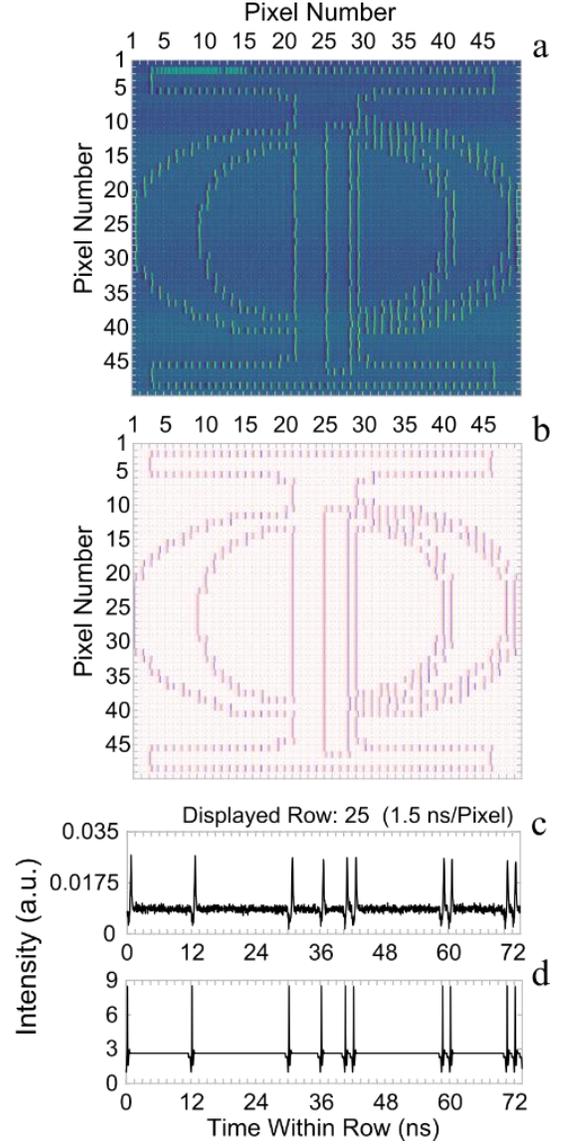

Fig 6. Experimentally measured (a) and theoretically calculated (b) responses from the VCSEL-neuron illustrating the successful performance of gradient magnitude detection in the 50x50 pixel image of the Institute of Photonics (IOP) logo. Similar to Fig. 5, (c) and (d) plot the measured and calculated time series at the output of the VCSEL-neuron for row 25 in (a) and (b), respectively.

## Acknowledgment

The authors would like to acknowledge Prof. A. Kemp and Prof. T. Ackemann from the University of Strathclyde for lending some of the equipment used in this work.

## REFERENCES

[1] N. Sebe, et al., (2005, Jun 3) *Machine learning in computer vision*. Springer Science & Business Media, 2006. Vol. 29.
[2] G. Litjens, et al., "A survey on deep learning in medical image analysis," *Medical Image Analysis*, vol. 42, pp. 60-88, 2017.
[3] J. Janai, et al., "Computer vision for autonomous vehicles: problems, datasets and state-of-the-art*.*" *arXiv preprint*, arXiv:1704.05519, 2017.
[4] O. Abdel-Hamid et al., "Applying convolutional neural networks concepts to hybrid NN-HMM model for speech recognition," *2012 IEEE international conference on Acoustics, speech and signal processing*, Kyoto 2012, pp. 4277-4280, 2012.
[5] A. Hannun et al., "Deep speech: scaling up end-to-end speech recognition," *arXiv preprint*, arXiv:1412.5567, 2014.
[6] K. Saravanan, S. Sasithra, "Review on classification based on artificial neural networks," *Int. J. Ambient Syst. Appl.*, vol. 2, no. 4, pp. 11-18, 2014.
[7] Y. LeCun, Y. Bengio, & G. Hinton, "Deep learning," *Nature*, vol. 521, pp. 436–444, 2015.
[8] M. El-Sayed, et al., "Automated edge detection using convolutional neural network," *Int. J. Adv. Comput. Sci. Appl.*, vol. 4, no. 10, pp. 11-17, 2013.
[9] D. Marr & E. Hildreth, "Theory of edge detection," *Proc. R. Soc. Lond.* B, vol. 207, pp. 187-217, 1980.
[10] H.C. Shin, et al., "Deep convolutional neural networks for computer-aided detection: CNN architectures, dataset characteristics and transfer learning," *IEEE Trans. Med. Imaging*, vol. 35, no. 5, 2016.
[11] C. Farabet, et al., "Learning hierarchical features for scene labeling," IEEE Trans. Pattern Anal. Mach. Intell., vol. 35, no.8, pp. 1915–1929, 2013.
[12] M. Rastegari, et al., "Xnor-net: imagenet classification using binary convolutional neural networks," *European Conf. Computer Vision 2016*, lecture Notes in Computer Science, *Springer, Cham*, vol. 9908, pp. 525-542, 2016.
[13] D.A.B. Miller, "Attojoule optoelectronics for low-energy information processing and communications," J. Lightwave Technol., vol. 35, no.3, pp. 346-396, 2017.
[14] A. Mehrabian, et al., "PCNNA: a photonic convolutional neural network accelerator," *2018 31st IEEE International System-on-Chip Conference*, Arlington, pp. 169-173, 2018.
[15] Y. Shen, et al., "Deep learning with coherent nanophotonic circuits," *Nature Photonics*, vol. 11, pp. 441-447, 2017.
[16] K. Shiflett, et al., "PIXEL: photonic neural network accelerator," *2020 IEEE Int. Symp. High Performance Computer Architecture,* San Diego, pp. 474-487, 2020.
[17] S. Xu, et al., "High-accuracy optical convolution unit architecture for convolutional neural networks by cascaded acousto-optical modulator arrays," *Opt. exp.*, vol. 27, no. 14, pp. 19778-19787, 2019.
[18] P. R. Prucnal, B. J. Shastri, and T. Ferreira de Lima, "Recent progress in semiconductor excitable lasers for photonic spike processing," *Adv. Opt. Photon.*, vol. 8, pp. 228–299, 2016.
[19] Y. Zhang, S. Xiang, X. Guo, A. Wen and Y. Hao, "Polarization-resolved and polarization-multiplexed spike encoding properties in photonic neuron based on VCSEL-SA," *Sci. Rep*., vol. 8, 16095, 2018.
[20] S. Xiang, A. J. Wen and W. Pan, "Emulation of spiking response and spiking frequency property in VCSEL-based photonic neuron," *IEEE Photonics J*., vol. 8, no. 5, 1504109, 2016.
[21] A. Hurtado, & J. Javaloyes, "Controllable spiking patterns in long-wavelength vertical cavity surface emitting lasers for neuromorphic photonics systems," *Appl. Phys. Lett.,* vol. 107, 241103, 2015.
[22] J. Robertson, T. Deng, J. Javaloyes, and A. Hurtado, "Controlled inhibition of spiking dynamics in VCSELs for neuromorphic photonics: Theory and experiments," *Opt. Lett.*, vol. 42, no. 8, pp. 1560–1563, 2017.
[23] Y. Zhang, et al., "All-optical inhibitory dynamics in photonic neuron based on polarization mode competition in a VCSEL with an embedded saturable absorber," *Opt. lett*., vol. 44, no. 7, pp. 1548-1551, 2019.
[24] T. Deng, J. Robertson, and A. Hurtado, "Controlled propagation of spiking dynamics in vertical-cavity surface-emitting lasers: Towards neuromorphic photonic networks," *IEEE J. Sel. Top. Quantum Electron.*, vol. 23, no. 6, 1800408, 2017.
[25] S. Xiang, et al., "Cascadable neuron-like spiking dynamics in coupled VCSELs subject to orthogonally polarized optical pulse injection," *IEEE J. Sel. Top. Quantum Electron*., vol. 23, no. 6, 1700207, 2017.
[26] T. Deng et al., "Stable propagation of inhibited spiking dynamics in vertical-cavity surface-emitting lasers for neuromorphic photonics networks," *IEEE Access*, vol. 6, pp. 67951–67958, 2018.
[27] J. Robertson, E. Wade, and A. Hurtado, "Electrically controlled neuron-like spiking regimes in vertical-cavity surface-emitting lasers at ultrafast rates," *IEEE J. Sel. Top. Quantum Electron.*, vol. 25, no. 6, 5100307, 2019.
[28] J. Robertson, et al., "Towards neuromorphic photonic networks of ultrafast spiking laser neurons," *IEEE J. Sel. Top. Quantum Electron.*, vol. 26, no. 1, 7700715, 2020.
[29] J. Robertson, et al., "Ultrafast optical integration and pattern classification for neuromorphic photonics based on spiking VCSEL neurons," *Sci. Rep.*, vol. 10, 6098, 2020.
[30] S. Xiang, Z. Ren, Y. Zhang, Z. Song, and Y. Hao, "All-optical neuromorphic XOR operation with inhibitory dynamics of a single photonic spiking neuron based on VCSEL-SA," *Opt. Lett.*, vol. 45, no. 5, pp. 1104-1107, 2020.
[31] J. Martin-Regalado, F. Prati, M. San Miguel, and N. B. Abraham, "Polarization properties of vertical-cavity surface-emitting lasers," *IEEE J. Quantum Electron.*, vol. 33, no. 5, pp. 765–783, 1997.
[32] H. Susanto, et al., "Spin-flip model of spin-polarized vertical-cavity surface-emitting lasers: asymptotic analysis, numerics, and experiments," *Phys. Rev. A,* vol. 92, 063838, 2015.
[33] S. Xiang et al., "STDP-based unsupervised spike pattern learning in a photonic spiking neural network with VCSELs and VCSOAs," *IEEE J. Sel. Top. Quantum Electron.*, vol. 25, no. 6, 1700109, 2019.
[34] S. Xiang, et al., "Numerical implementation of wavelength-dependent photonic spike timing dependent plasticity based on VCSOA," *IEEE J. Quantum Electron*., vol. 54, no. 6, 8100107, 2018.

**Joshua Robertson** was born in Glasgow, U.K., in 1994. He received the MPhys degree in physics with specialization in photonics, in 2017, from the University of Strathclyde, Glasgow, U.K., where he is currently working toward the Ph.D. degree with the Institute of Photonics, focusing on neuromorphic photonic systems with lasers.

**Yahui Zhang** was born in Hebei Province, China, in 1993. She is working toward the Ph.D. degree with Xidian University, Xi'an, China, and is currently spending a research year collaborating with colleagues at the University of Strathclyde, Glasgow, U.K. Her research interests include neuromorphic photonic systems, brain-inspired information processing and random number generators.

**Matěj Hejda** was born in Plzeň, Czech Republic in 1993. He received his master's Ing. degree from Technical University of Liberec in collaboration with ICFO, Barcelona in 2019. He is currently working toward the Ph.D. at the Institute of Photonics, University of Strathclyde, Glasgow, U.K., studying neuromorphic optoelectronic and nanophotonic systems.

**Andrew Adair** was born in Glasgow, U.K. in 1998. He is currently working towards his MPhys degree at the University of Strathclyde, Glasgow, U.K., where he recently completed his experimental project in the neuromorphic photonics group.

**Julián Bueno** was born in Palma de Mallorca, Spain, in 1988. He received the B.Sc. degree in physics from the Universitat de les Illes Balears (UIB), in 2013, and the M.Sc. degree in complex systems from the Institute for Cross-Disciplinary Physics and Complex Systems IFISC (CSIC-UIB), in 2014. He received the Ph.D. degree with UIB, in 2019, and has recently joined as a Research Associate with the Institute of Photonics, University of Strathclyde, Glasgow, U.K., working in the "BRAIN LASER" project funded by ONRG to develop novel neuromorphic photonic systems and networks with laser-based optical neurons. His research interests include nonlinear photonics, complex dynamics, and the implementations of neural networks in photonic hardware.

**Shuiying Xiang** was born in Ji'an, China, in 1986. She received the Ph.D. degree from Southwest Jiaotong University, Chengdu, China, in 2013. She is currently a Professor with the State Key Laboratory of Integrated Service Networks, Xidian University, Xi'an, China. She is the author or coauthor of more than 80 research papers. Her research interests include neuromorphic photonic systems, brain-inspired information processing, spiking neural network, vertical cavity surface-emitting lasers, and semiconductor lasers dynamics.

**Antonio Hurtado** received the Ph.D. degree from Universidad Politécnica de Madrid (UPM), Madrid, Spain, in December 2006. He has pursued an international photonics research career working in the UK (Universities of Essex and Strathclyde), USA (University of New Mexico), and Spain (UPM). He received two Marie Curie Fellowships from the European Commission: Projects ISLAS (2009–2011) and NINFA (2011–2014). In 2014, he was awarded a Chancellor's Fellowship by the University of Strathclyde following which he was appointed as a Lecturer (Assistant Professor) with the Strathclyde's Institute of Photonics (Physics Department). He was promoted to Senior Lecturer (Associate Professor) in 2018. He has been involved in multiple research projected funded by UK, US and European Union (EU) funding agencies. His current research interests are but not limited to neuromorphic photonics, nanophotonics, laser nonlinear dynamics, nanolaser systems, and hybrid nanofabrication.